\documentclass{article}
\usepackage{template}

\usepackage[utf8]{inputenc} 
\usepackage[T1]{fontenc}    
\usepackage{booktabs}       
\usepackage{amsfonts}       
\usepackage{amsmath}
\usepackage{nicefrac}       
\usepackage{microtype}      
\usepackage{graphicx}
\usepackage{natbib}
\usepackage{url}

\usepackage{xcolor, colortbl}
\usepackage{subfiles}
\usepackage{multirow}
\usepackage{soul}
\usepackage{enumitem}

\definecolor{lightpurple}{RGB}{217,210,233}
\definecolor{lightteal}{RGB}{208,223,226}
\definecolor{lightpink}{HTML}{E9D3D3}
\definecolor{darkgreen}{rgb}{0.0, 0.5, 0.0}
\definecolor{citeblue}{rgb}{0, 0, 123}

\usepackage[colorlinks=true,
linkcolor=citeblue
,citecolor=citeblue,
,filecolor=red
,urlcolor=citeblue
,menucolor=red
,runcolor=red
,breaklinks=true]{hyperref}

\renewcommand{\rhead}{Goyal et al.}

\title{Causal Effect of Group Diversity\\ on Redundancy and Coverage in Peer-Reviewing}

\author{
 Navita Goyal$^1$, Ivan Stelmakh$^2$, Nihar Shah$^3$, Hal Daumé III$^1$ \\  
 $^1$University of Maryland, $^2$New Economic School, $^3$Carnegie Mellon University \\
 \texttt{navita@umd.edu, istelmakh@nes.ru, nihars@cs.cmu.edu, hal3@umd.edu}
}

\AtBeginDocument{%
    \renewcommand\sectionautorefname{\S\kern-0.2em}
    \renewcommand\subsectionautorefname{\S\kern-0.2em}
    \renewcommand\subsubsectionautorefname{\S\kern-0.2em}
}

\begin{document}
\maketitle

\setlist{leftmargin=1em,itemsep=5pt,parsep=0pt,topsep=2pt}

\begin{abstract}
  A large host of scientific journals and conferences solicit peer reviews from multiple reviewers for the same submission, aiming to gather a broader range of perspectives and mitigate individual biases. In this work, we reflect on the role of diversity in the slate of reviewers assigned to evaluate a submitted paper as a factor in diversifying perspectives and improving the utility of the peer-review process. We propose two measures for assessing review utility: review coverage---reviews should cover most contents of the paper---and review redundancy---reviews should add information not already present in other reviews. We hypothesize that reviews from diverse reviewers will exhibit high coverage and low redundancy. We conduct a causal study of different measures of reviewer diversity on review coverage and redundancy using observational data from a peer-reviewed conference with approximately 5,000 submitted papers. Our study reveals disparate effects of different diversity measures on review coverage and redundancy. Our study finds that assigning a group of reviewers that are topically diverse, have different seniority levels, or have distinct publication networks leads to broader coverage of the paper or review criteria, but we find no evidence of an increase in coverage for reviewer slates with reviewers from diverse organizations or geographical locations. Reviewers from different organizations, seniority levels, topics, or publications networks (all except geographical diversity) lead to a decrease in redundancy in reviews. Furthermore, publication network-based diversity alone also helps bring in varying perspectives (that is, low redundancy), even within specific review criteria. Our study adopts a group decision-making perspective for reviewer assignments in peer review and suggests dimensions of diversity that can help guide the reviewer assignment process.
\end{abstract}



\section{Introduction}

Peer-reviewing is an integral part of the research process \citep{jefferson2002effects, ware2008peer, shah2022overview}. The peer-review process typically involves collating judgments from multiple reviewers to assess a submission. The group of reviewers may directly (through discussion) or indirectly (in the form of written reviews) influence the final decision, typically made by the members of the editorial committee at the journal or conference. 
Having multiple reviewers look at one submission serves various purposes, such as bringing in different perspectives, ensuring unbiased opinion, providing broad feedback, and reducing noise in the reviewing process \citep{grimaldo2013simulation, bianchi2015three}. Drawing on this, we posit two desiderata of peer-reviewing: \textit{(1) reviews should collectively cover most, if not all, aspects of the submitted paper and the review criteria}, and \textit{(2) each review should contribute additional insights and perspectives}.

Desideratum 1 underscores the importance of achieving comprehensive \textit{coverage} in reviews. While it may be unrealistic to expect for every reviewer to possess expertise in all areas or evaluate every aspect of a paper, we aim for reviewers to collectively provide a comprehensive assessment, enabling decision-makers to make a well-informed decision~\citep{porter1985peerinterdisciplinary,  Laudel2006Interdisciplinary, shah2022overview}. We delineate two notions of coverage: paper coverage, which pertains to the coverage of various facets of the submitted paper, and type coverage, which focuses on the coverage of different review criteria. 
Desideratum 2 underscores the importance of having a diverse perspective in reviews. 
Research in \textit{agreement in peer review} reflects that ``diversity of opinion among referees may be desirable and beneficial, bringing new and broader perspectives to the review process'' \citep{chubin1990peerless, hargens1990neglected, lee2013bias}. 
We operationalize the diversity of perspectives in reviews as low redundancy, positing that reviews with low redundancy introduce different perspectives for a more informed and unbiased assessment. 
Reviewing requires a lot of time and effort by reviewers---time and effort spent \textit{not} doing research. A high redundancy suggests duplication, and hence, wasted effort. Moreover, since peer-reviewing also serves as a means for authors to obtain feedback for improving their submitted papers, redundant reviews are not particularly helpful to the authors either. 
Unlike coverage, which assesses the extent to which different facets of the paper or review criteria are addressed across the reviews, redundancy examines the subjective variations in the evaluation for each facet between the reviews. Thus, we use coverage and redundancy as the two metrics to assess the effectiveness of the peer-review process from a group decision-making perspective. 

Review coverage and redundancy are properties of the reviews, whereas the crucial lever at the disposal of journal editors or program chairs lies in the recruitment and assignment of reviewers. 
Since review coverage and redundancy can only be assessed after the reviews are written, we explore whether reviewer profiles---such as affiliation, seniority, and publication network---available during reviewer assignments can serve as proxies. 
Our work connects these reviewer attributes with review outcomes, showing how specific reviewer assignment strategies can shape these characteristics of reviews. 
More specifically, we consider the diversity of reviewer assignments as a potential way to introduce broader coverage and perspectives in the reviews~\citep{jackson1995understanding, olbrecht2010panelpeerreview, shah2022overview}. 

Despite the potential advantages of incorporating diversity in the review assignment process, asserting multiple dimensions of diversity may constrain the assignment too much, making it challenging to find appropriate reviewers for a large pool of submitted papers. For instance, we can consider diversity along various dimensions---assign reviewers from different organizations, countries, seniority levels, and so forth and it is not clear how such different dimensions of reviewer diversity affect the peer-review process. Thus, there is a need for a systematic study of the effect of the different dimensions of reviewer diversity on the peer-review process. To this end, we conduct a causal study of the effect of reviewer slate diversity on the coverage and redundancy of the reviews. 

Conducting a large-scale randomized study to evaluate these effects is difficult, especially in the context of scientific peer-reviewing, and sometimes may lead to ethical concerns. 
Therefore, we study the potential effects of diversity in reviewer assignment on review utility using review data from a past conference, spanning close to $5000$ papers.\footnote{In Computer Science, conferences are typically considered at par with journals, and conference publications undergo a comprehensive peer-review of the whole paper, rather than just abstracts. The review data in our study comes from International Conference on Machine Learning (ICML) 2020, which is one of the most prestigious machine learning conferences with an acceptance rate of $21.8\%$.
} 
As is typical, observational peer-review data are plagued with multiple confounders, such as reviewers' expertise in evaluating a submitted paper, reviewers' own profile, or the topic and content of the paper. 
For instance, assigning topically diverse reviewers to a paper may increase the diversity of the reviewer slate, but the reviewers may have less overall expertise in the paper's topic, which is not desirable. 
Despite the potential benefits of diverse reviewer slate, it is still crucial to ensure that reviewers' expertise is not compromised. Therefore, any intervention in reviewer assignment should carefully consider factors such as reviewer-paper expertise and other relevant confounders. 
We propose a causal framework that accounts for these confounding factors to accurately estimate the causal impact of the variables of interest (diversity) on the outcome (coverage and redundancy). 

Our research advocates for a systematic study of how the diversity within the group of reviewers assigned to a submitted paper impacts the coverage and redundancy in reviews in a peer-review setup. The significant immediate consequences of the peer-review process, ranging from employment considerations to funding allocations, underscore the importance of a methodical approach to making major changes in the peer-review system. 
Moreover, given the extensive community involvement inherent in peer review, empirical evidence supporting alterations in the peer-review process not only offers early indications of successes and failures but also garners support from various stakeholders, including authors, reviewers, and the editorial boards of peer-reviewed journals and conferences. 
By fostering empirical investigation into changes in the reviewer assignment process, we can ensure that modifications are justified and grounded in data, ultimately enhancing both the transparency and effectiveness of the peer-review process.

\section{Related Work}

In this section, we discuss past literature that is related to the topic of this paper.  

\paragraph{Diversity in collaborations and group decision-making}
The role of diversity in group problem-solving, collaborations, and decision-making has been extensively studied in various fields, including organizational behavior, sociology, economics, and more. A seminal work by \citet{page2008perspectives} argues that diversity in perspectives enhances innovation, scientific understanding, and problem solving.   
\citet{hong2004problemsolving} further present a framework for modeling diverse teams and conclude that the diversity of an agent's problem-solving approach relative to other problem solvers is an important predictor of their value and may even be more relevant than their individual ability. \citet{reynolds2017teams} find that cognitively diverse teams solve problems faster. \citet{bantel1989top} study 199 banks and find that banks managed by more diverse teams with respect to their functional areas of expertise are more innovative. 

Beyond diversity in functional expertise, various other works highlight the potential benefits of demographic diversity in enhancing group decision-making. \citet{sommers2006racial} finds that racial diversity in juries leads to increased information sharing. Further, \citet{levine2014ethnic} find that ethnic diversity in markets deflates price bubbles, whereas homogenous markets result in overpricing. 
Several studies on organizations and team effectiveness find that the diversity of organizations, such as gender diversity, age diversity, and more, is directly linked to their performance \citep{hoffmann1962differences, jackson1995understanding}. Specifically in the context of scientific research, \citet{Hofstra2020diversityparadox} find that diversity in demographics leads to innovation. In all, this prior evidence suggests that ensuring diversity in reviewer slates could improve the utility of reviews in
the peer-review process.

\paragraph{Diversity in peer-reviewing} Peer-review plays an integral role in assessing the quality of scientific research \citep{jefferson2002effects, ware2008peer, shah2022overview}, and thus, introducing diversity in the peer-reviewing mechanism can aid decision-making. 
\citet{olbrecht2010panelpeerreview} argue that homogeneity in panel peer review, for instance, panels made up of scientists belonging to the same organization, can lead to faulty decision-making, concluding that group members should
be as diverse as possible as ``group diversity leads to different points of view and perspectives in the discussion''. 
\citet{shah2022overview} argues that diversity in reviewers evaluating a submission can help in ensuring higher coverage in evaluating interdisciplinary results. 
Further, \citet{jecmen2020mitigating} argue that diversity among reviewers can also mitigate collusion rings in the peer-reviewing process (wherein reviewers within the same network collude to review each other’s work for unfair advantage). They develop algorithms that assign reviewers to papers with some randomness, that are now deployed by many conferences. Such randomness can in turn also allow for better measurements of counterfactual outcomes~\citep{saveski2023counterfactual}. 

\citet{AAAI2024papermatching} propose maximizing geographical diversity and co-authorship distance between reviewers assigned to a submission. \citet{kuznetsov2024whatcannlpdo} also discuss diversity constraints in reviewer assignments in large peer-reviewed conferences. 
However, these papers do not empirically study the potential impact of these diversity constraints on the utility of the peer-reviewing process. Our work aims to fill this gap by systematically studying how different axes of diversity can potentially help in peer-reviewing.

Empirically, \citet{murray2018international} study peer-reviewing in biosciences journals and find that editors and peer reviewers tend to favor manuscripts from authors of the same gender and from the same country, thus concluding that gender and international diversity in reviewers can improve equity in peer review. 
\citet{dumlao2023geographical} 
also find evidence of geographical homophily in peer-reviewed journals, wherein, reviewers favor work from authors from the same country, suggesting reviewer diversification and double-anonymization as potential ways to mitigate geographical representation bias.
We deviate from these studies by extending beyond acceptance decisions and scores to examine how reviewer diversity can contribute to broader, more unbiased reviews, focusing specifically on the review text. 

\citet{schouten2023broaderpanels} study proposal reviews in health research and find that ``more diverse panels [healthcare professionals, academics, patients, and policymakers] result in a wider range of arguments, largely for the benefit of arguments related to societal relevance and impact''. 
However, this work focuses solely on diversity in the background among panelists (such as healthcare professionals, academics, patients, and policymakers). We extend this to different axes of diversity, such as organizational and geographical diversity among reviewer panels. Secondly, in their experimental setting, the panelists discussed the proposals in peer-review meetings, where different panelists actively interact and potentially affect other panelists explicitly. In contrast, our study investigates an offline peer-review setting where the reviewers do not interact among themselves or observe each other's reviews, except until after submitting their reviews.


\section{Review coverage and redundancy}
Peer-reviewing is a time-intensive and laborious task, which, if done right, has the potential for great value. 
At its best, the peer-review process can serve two goals: informing decision makers (e.g., action editors) about the strengths and weaknesses of a submitted paper to assess readiness for publication, and providing feedback to the authors about potential improvements to strengthen the work. Typically, peer-reviewing entails multiple reviewers assessing each submitted paper, which helps bring in a broader perspective and reduce bias. However, previous work predominantly discusses the utility of reviews individually \citep{jefferson2002quality, Xiong2011Helpfulness, hua2019argument, yuan2022automated}. We consider two measures of review utility at a group level: review coverage and review redundancy with the following desiderata:

\paragraph{Desideratum 1: Reviews should collectively cover most aspects of the paper.} Review coverage aims to capture the comprehensiveness of the reviews. 
We can not expect each individual reviewer to be an expert in everything or assess every aspect of the paper. 
However, if reviewers can collectively assess the paper comprehensively, then decision-makers can make an informed decision, and authors can have broad feedback \citep{porter1985peerinterdisciplinary, Laudel2006Interdisciplinary, shah2022overview}. 
We consider two notions of coverage: 
\begin{itemize}
    \item \textit{Paper coverage:} Paper coverage pertains to the coverage of various aspects of the submitted paper. For instance, a paper may contain both theoretical and empirical contributions. Paper coverage assesses how well these facets are addressed across the reviews.
    \item \textit{Type coverage:} Type coverage focuses on the coverage of different review criteria. For instance, submissions are evaluated along multiple criteria such as novelty, clarity, correctness, and replicability, among others. Type coverage assesses how well these different assessment criteria are addressed across the reviews. 
\end{itemize}
We describe our measures of review coverage in \autoref{subsec:coverage}. 

\paragraph{Desideratum 2: Each review should contribute additional insights and perspectives.}
From a group decision-making perspective, the maximal utility of the peer-review process is to have reviews that add information not already present in other reviews \citep{hargens1990neglected}. 
We operationalize this as redundancy, positing that reviews with low redundancy introduce different perspectives for a more informed and unbiased assessment. 
It's important to note that lower redundancy may naturally accompany high coverage; reviews with broader coverage are more likely to result from different reviews assessing different parts of the paper or different review criteria, thus resulting in lower redundancy.
However, redundancy focuses specifically on assessing (dis)agreements in the subjective evaluation of specific facets of the submitted paper or the review criteria. 
For instance, a pair of reviews, both addressing the ``empirical results'' of a paper from the lens of ``correctness'', may disagree on the validity of the baseline choices. In this case, a pair of reviews that agree on the baseline choice would have similar coverage as a pair of reviews that do not, but a higher redundancy. 
Although such disagreements might complicate decision-making, the inclusion of diverse perspectives in the review process can sufficiently highlight critical considerations for making informed decisions. It is worth noting that submissions with apparent strengths or flaws may naturally exhibit redundancy in reviews. However, as our analysis controls for the paper, non-redundancy between reviews may indicate novel arguments and viewpoints. We describe our measures of review redundancy in \autoref{subsec:redundancy}.

Coverage and redundancy can be viewed as complementary metrics that offer distinct perspectives on the quality and effectiveness of peer reviews. Coverage serves as a surface-level metric, gauging the breadth of the review process by assessing how extensively different aspects of the paper or evaluation criteria are addressed. On the other hand, redundancy digs deeper into the content of the reviews, evaluating the subjective differences in the assessment of different facets of the paper or review criteria.

\section{Diversity in reviewer slates}\label{sec:hypotheses}

The diversity of groups plays an important role in decision-making. Research in teams and collaboration suggests that diversity enhances decision-making by increasing the information and perspectives in the processes \citep{jackson1995understanding}. This prior evidence suggests that ensuring diversity in reviewer slates could improve the utility of reviews in the peer-review process. In this context, diversity has been discussed as a means to achieve higher coverage when evaluating interdisciplinary research \citep{shah2022overview}. Furthermore, a diversity of viewpoints can help reduce bias in the reviews, as reviewers from similar research backgrounds may have biased expectations, putting research outside such clusters at a disadvantage. 

The research, both in group decision-making and peer-reviewing, approaches diversity along various axes: gender, race, organization, nationality, area of expertise, years of experience, etc. \citep{jackson1995understanding}. 
It is not clear which axes of diversity are potentially important in improving the efficacy of the peer-review process. We investigate this by considering reviewer diversity on the following axes: 
\begin{itemize}
    \item \textit{Organizational}: whether the reviewers belong to the same organization, 
    \item \textit{Geographical}: whether the reviewers are from the same geographical location, 
    \item \textit{Co-authorship}: whether the reviewers have co-authored papers together or have co-authors in common,\footnote{Common co-authors condition targets reviewers who do not publish together, but may have a strong common influence, for instance, students with the same advisor.}
    \item \textit{Topical}: whether the reviewers work on the same topics, and
    \item  \textit{Seniority}: whether the reviewers have the same seniority level.
\end{itemize}
We describe how we obtain these diversity measures in \autoref{subsec:diversity}. 
In this work, we develop a causal framework to formally assess the effects of different measures of reviewer diversity on different measures of review coverage and redundancy. Specifically, we test the following hypotheses: 
\begin{itemize}
     \item \textit{Hypothesis 1:} 
    Diverse reviewer slates have \textbf{higher coverage} than non-diverse reviewer slates.
    \item \textit{Hypothesis 2:} 
    Diverse reviewer slates have \textbf{lower redundancy} than non-diverse reviewer slates.
\end{itemize}

\section{Method}\label{sec:causalframework}

In this section, we outline the method used in our work to investigate the hypotheses described in \autoref{sec:hypotheses}. Reiterating, the goal is to measure the causal effects of assigning diverse reviewers to a submitted paper on the coverage and redundancy in the reviews. We first detail the dataset and the setting in which this data was collected in \autoref{subsec:data}. Next, we outline our outcome measures, confounders, and treatment in \autoref{subsec:outcomes}, \autoref{subsec:confounders}, and \autoref{subsec:diversity}. Lastly, we describe two effect estimation approaches putting these all together in \autoref{subsec:approach}.

\subsection{Experimental setting} \label{subsec:data}
We first detail the experimental setting of our work. Our experiments are based on peer-review data\footnote{The University of Maryland Institutional Review Board has approved the use of the data for research purposes.} from a popular machine learning conference---International Conference on Machine Learning (ICML 2020).\footnote{\url{https://icml.cc/}} ICML 2020 comprises of a total of $4,991$ submitted papers and a pool of $3,637$ reviewers, with each paper being evaluated by $2$ to $4$ reviewers ($2: 4.1\%, 3: 71.9\%, 4: 23.6\%$). The peer-review process at ICML is double-blind and involves an automated assignment system that matches submitted papers with reviewers based on their prior publications. The assigned reviewers are tasked with providing a numerical review score ranging from $1$ (Strong reject) to $10$ (Strong accept), along with a comprehensive review text, with separate fields for adding the submitted paper's summary, strengths, weaknesses, and any additional comments.
Additionally, reviewers specify their expertise on the paper and their confidence in their assessment.  

Following the initial review, authors have the opportunity to engage in a rebuttal phase, during which they receive and respond to the reviewers' feedback. Reviewers may then engage in discussions regarding the reviews and the authors' responses. Reviewers then have a chance to revise their reviews and scores based on the authors' responses and any ensuing discussions. The final acceptance decision is made by meta-reviewers and area chairs based on the collective reviews and scores. 

To prevent inter-review effect leakage during the rebuttal and discussion phases, our analysis focuses solely on pre-rebuttal reviews. We aggregate the summary, merits, demerits, and comments into a single review text. This avoids any inconsistencies in how the review is separated across different elements. For example, some reviewers add all their comments in the summary and then include ``refer to the summary for merits/demerits'', etc. We consider textual reviews (hereafter, reviews) instead of review scores, as scores are only shallow indicators of reviewers' opinions---multitudes of factors may result in the same score. Textual reviews, on the other hand, reveal detailed dimensions considered by different reviewers, offering a comprehensive view of the effects of diversity on the peer-review process \citep{Lee2012Kuhnian}. 

Lastly, in addition to submitted paper and review data, the ICML 2020 data also contains reviewer profile information, such as their email ID, organization, location (country), Semantic Scholar, and Google Scholar profiles, some of which are missing for some reviewers---organization ($4.8\%$), location ($13.3\%$), Semantic Scholar profile ($51.0\%$), Google Scholar profile ($17.5\%$). We extract reviewers' recently published papers and h-index from their Google Scholar profiles.\footnote{The Google Scholar profile is based on the data scraped in 2021.} Before conducting our analysis, we anonymize reviewers and all associated profile information. 

\subsection{Outcomes}
\label{subsec:outcomes}
As mentioned above, we use coverage and redundancy between the textual reviews written by different reviews for the same submitted paper. In this section, we detail the different measures of review coverage and review redundancy used in our study. 

\subsubsection{\textbf{Review Coverage}} \label{subsec:coverage}
We assess review coverage using two measures of coverage: type coverage (aspects and arguments) and paper coverage (lexical and semantic). 
 
\paragraph{\textbf{Type coverage}.} We evaluate type coverage using different typologies of reviews discussed in previous work: (1) the \textit{aspect} of the submitted paper being commented upon, namely, Summary, Motivation/Impact, Originality, Soundness/Correctness, Substance, Replicability, Meaningful Comparison, and Clarity \citep{yuan2022automated} and (2) the type of \textit{argument} being presented, namely, Evaluation, Fact, Request, Reference, and Quote \citep{hua2019argument}. We measure type coverage as the number of aspect or argument types that are covered across the pair of reviews. 

We first annotate the sentences in each review with aspect and argument types using an automated model that takes sentences from the review text as input and predicts their type. This annotation process employs a transformer-based sequence classification model, specifically DistilBERT \citep{sanh2019distilbert}. The model is initialized with pre-trained parameters learned from large unsupervised text corpora and then fine-tuned for aspect and argument classification using datasets labeled with the respective categories. The models achieve a held-out test accuracy of $90.11\%$ and $82.96\%$ on their respective datasets. 

\paragraph{\textbf{Paper coverage}.} In paper coverage, we measure the overlap between the reviews and the submitted paper's abstract in two ways: lexical and semantic. \textit{Lexical coverage} evaluates the extent to which lexical units of the paper, as conveyed by its abstract, are covered across the two reviews. We quantify lexical coverage as the ratio of unique $n$-grams present in the paper's abstract that are also present in at least one review. 

\textit{Semantic coverage} evaluates the extent to which the reviews span the semantic units conveyed in the paper's abstract. To quantify semantic coverage, we leverage a high-dimensional semantic representation of the review text and abstract, specifically using the sentence-BERT model\citep{reimers2019sbert}. For each sentence in the paper abstract, we compute its coverage score as the semantic similarity with the most similar review sentence across the pair of reviews. 

\subsubsection{\textbf{Review Redundancy}} \label{subsec:redundancy}
We assess review redundancy using three measures: lexical redundancy, semantic redundancy, and weighted semantic redundancy. 

\textit{Lexical redundancy} evaluates the overlap in lexical units between the two reviews. We quantify lexical redundancy as the intersection of $n$-grams in the reviews, with $n$ ranging from $1$ to $3$.

\textit{Semantic redundancy} evaluates the overlap in the semantic units in the two reviews. We quantify semantic redundancy as the cosine similarity between the high-dimensional representation of the review sentences. Specifically, we consider the aggregate similarity of sentences in one review (\textit{reference}) with the most similar sentence in the other review (\textit{target}). To ensure symmetry in the measure, we consider each review as the reference review in turn. 

\textit{Weighted semantic redundancy} evaluates the overlap in the semantic units in the two reviews weighted by their aspect or argument type. The idea behind this measure is that semantically similar reviews within a specific aspect or argument type would suggest higher redundancy. Therefore, to adjust for the similarity of the review types, we weigh the cosine similarity of the semantic representation with the similarity between their type classification. 

Please refer to \autoref{appendix:formulas} for formulas of the outcome measures. To ensure a consistent estimation of effect sizes, we normalize all outcome measures to fall within the range of $0$ and $1$ (details in \autoref{appendix:postprocessing}). We additionally also validate our automated outcome measures with a small human annotation (details in \autoref{appendix:human}). 

\subsection{Confounding factors}
\label{subsec:confounders}
\begin{figure}[t]
    \centering
    \includegraphics[width=0.5\linewidth]{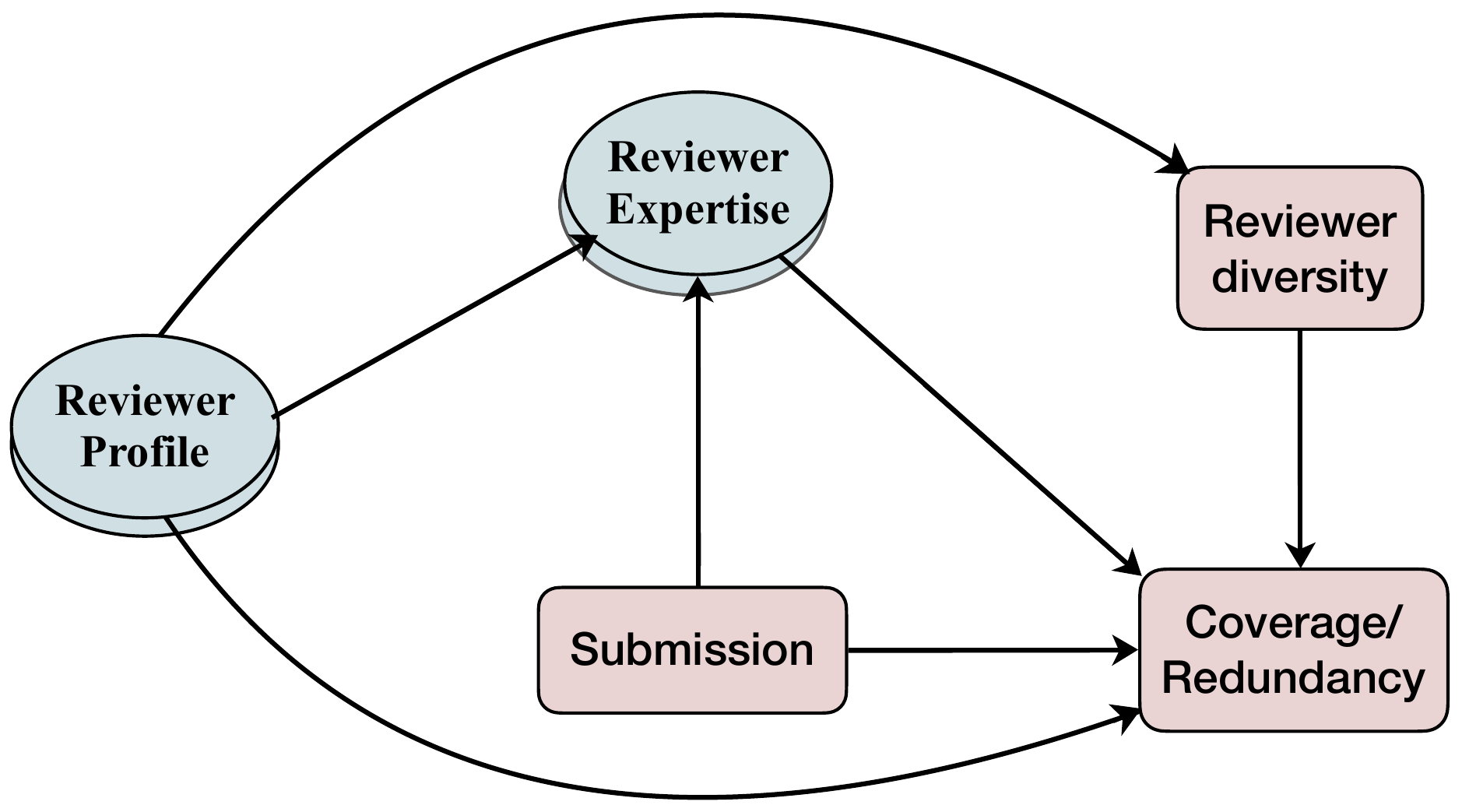}
    \caption{Causal Diagram consisting of reviewer-specific features (oval nodes)---reviewer profile and reviewer expertise with respect to the submitted paper---and slate features (square nodes)---submission, reviewer slate diversity, and review coverage and redundancy. Reviewer profiles dictate reviewer expertise with respect to the submitted paper and the diversity of the reviewer slate. The combination of reviewer profile, expertise, and submitted paper determines the reviews, which in turn affect review coverage and redundancy.
    }
    \label{fig:causal_diag_simplified}
\end{figure}

As discussed previously, we are interested in studying whether assigning a diverse slate of reviewers to a submitted paper can help enhance the quality of reviews (in terms of review coverage and redundancy) using observational data from ICML 2020. As is typical, such observational data is fraught with multiple confounding factors. Thus, to accurately estimate these causal effects, we need to account for these confounding factors. 
We consider the following confounding factors in our analyses:
\begin{itemize}
    \item(C1) \textit{Submitted paper content.} The content of the submitted paper plays a major role in shaping reviews and, consequently, coverage and redundancy measures. In parallel, the paper content also determines which reviewer gets assigned to the paper. 
    For instance, interdisciplinary papers may naturally get diverse set of reviewers, but these papers may be likely to have high variance (low redundancy) in reviews. This may erroneously amplify the potential impact of diversity on review coverage and redundancy or miss an effect where one exists. 
    Thus, paper content, such as topic and quality, which affects both the treatment (reviewer assignment) and the outcome (review coverage and redundancy), needs to be accounted for in the causal analyses. 
    \item(C2) \textit{Reviewer expertise.} The reviewer's expertise with respect to a submitted paper simultaneously affects both the reviewer assignment and the reviews and, subsequently, the overall coverage and redundancy. For instance, having one or more reviewers with low expertise may introduce diversity in the reviewer slate but can also result in poor reviews with low coverage. Thus, we need to account for reviewer expertise in our analyses to avoid missing any potential effect of diversity in improving review coverage or redundancy. 
    \item(C3) \textit{Reviewer profile.} Reviewer profiles, including, their affiliation, location, area of research, and seniority, are all responsible for shaping reviewers' preferences, reviewing style, and evaluation strictness. However, reviewer pools often tend to be non-uniform, with variations in the representation of different organizations, geographical locations, topics, or seniority levels. Consequently, non-diverse reviewer slates may disproportionately include reviewers from specific locations, organizations, etc. As a result, even in the absence of any effect of reviewer slate diversity, there could be differences in review coverage and redundancy between diverse and non-diverse reviewers. Thus, these confounding factors could potentially bias analyses and must be carefully accounted for.  
\end{itemize}
\autoref{fig:causal_diag_simplified} shows the causal graph indicating the relationship between the treatment (diversity of reviewer slates), outcome (review coverage and redundancy), and the various confounding factors (submitted paper, reviewer expertise with respect to the submission, and the reviewer profile).

We describe how we control for confounder C1 in the next section. For confounder C2, we denote the reviewer $r$'s expertise score for the submitted paper $S$ as $E(r, S)$. 
We assess this expertise as the overlap between the text of the submitted paper and the reviewer's past papers. 
Due to the high volume of submitted papers, large conferences typically employ an automated reviewer assignment process. ICML 2020 performed reviewer assignments using The Toronto paper matching system (TPMS) \citep{charlin2013toronto}. 
This system computes reviewers' expertise scores for each submitted paper, known as the TPMS score, as the similarity between the set of reviewers' published papers and the submitted paper's text. This is followed by a constrained optimization problem aiming to maximize the joint TPMS scores for the reviewer slate assigned to all submitted papers while ensuring that each submitted paper is reviewed by the target number of reviewers and each reviewer has the desired reviewing load. 
In our analysis, we use these TPMS score between the reviewer and the submitted paper as the reviewer expertise score $E(r, S)$. 
This information is available in the ICML 2020 data. We do not use reviewers' self-reported measure of expertise due to potential issues with calibration in self-assessment. 

For confounder C3, we consider reviewer affiliation, location, area of research, seniority, and co-authorship information. We denote these dimensions as $d$, such that $d\in \{$organization, geographical location, seniority, topic, and co-authorship$\}$. We represent reviewer $r$'s profile vector along dimension $d$ as $\mathbf{d}(r)$. 
We obtain reviewers' affiliations and geographical locations directly from the review data. We represent a reviewer's organization as a $945$-dimensional vector covering all organizations in the dataset, with $1$'s for the reviewer's marked affiliation(s). Similarly, we represent a reviewer's geographic location as a $12$-dimensional one-hot vector based on $12$ geographical regions obtained by grouping countries (e.g., India $\rightarrow$ South Asia). This grouping ensures that there are multiple reviewers in each region. 
We obtain reviewers' seniority based on their h-index from Google Scholar profiles. We represent a reviewer's seniority using a $2$-dimensional one-hot vector, where reviewers with an h-index above the median value of 22 are categorized as senior, and those below as non-senior. For co-authorship information, we use reviewers' listed publications and corresponding co-authors from their Google Scholar profiles. Subsequently, we represent a reviewer's co-author vector as a $39291$ dimensional vector, with $1$'s for the reviewer's listed co-authors.\footnote{Due to the sparsity of this feature, it is excluded during the fixed effect estimation process.} 
For reviewers' topics, we use abstracts from their previously authored papers available on their Google Scholar profiles. We apply a Latent Dirichlet Allocation-based topic model \citep{blei2001lda} on these abstracts, resulting in a $10$-dimensional topic vector for each reviewer (refer to the \autoref{appendix:topic} for further details).

\subsection{Treatment}\label{subsec:diversity}
The treatment in our analyses is the diversity of reviewers assigned to a submitted paper. 
Consider a pair of reviewers $(r_1, r_2)$ assigned to a submitted paper $S$. Even though there are frequently more than two reviewers assigned to each paper, we consider the diversity of a pair of reviewers at a time to allow us to control for submission across diverse and non-diverse reviewer pairs (more details to follow in \autoref{subsec:approach}). 
We denote the diversity of the pair of reviewers along the dimension $d$ as $\delta_d(r_1, r_2)$. We consider reviewer diversity along $5$ dimensions: 
$d\in D=\{$organization, geographical location, seniority, topic, and co-authorship$\}$. The dimensions considered for the diversity of the two reviewers are the same as those considered in the individual reviewer's profile (confounder C3). Subsequently, we say that the reviewer diversity $\delta_d(r_1, r_2)=1$ if the pair of reviewers $r_1$ and $r_2$ are diverse and $\delta_d(r_1, r_2)=-1$ if the pair of reviewers $r_1$ and $r_2$ are non-diverse along dimension $d$. For instance, if the reviewers $r_1$ and $r_2$ have the same geographical location, then their $\delta_{\text{geographical}}(r_1, r_2)$ equals $-1$.

For organizational, geographical, topical, and seniority-based diversity, we use the respective profile vectors $\mathbf{d}(r_1)$ and $\mathbf{d}(r_2)$ for the two reviewers. In the case of organizational, geographical, and seniority-based diversity, we determine the value of $\delta_d(r_1, r_2)$ as $-1$ if $\mathbf{d}(r_1)$ and $\mathbf{d}(r_2)$ have $1$'s in the same position, indicating that the two reviewers share the same organization, geographical location, or seniority-level, and $1$ otherwise. 

Since topics are not represented by binary vectors, we quantify reviewer diversity along the topic dimension using the dot product between the topic vectors of the two reviewers. This dot product yields a topical similarity score between $0$ and $1$. Finally, to compute $\delta_{\text{topic}}(r_1, r_2)$, we binarize this topical similarity using the median value of $0.6472$. Hence, we consider reviewers having a topical similarity of $\geq$$0.6472$ as non-diverse (that is, $\delta_{\text{topic}}(r_1, r_2)=-1$), and those with a topical similarity of $<$$0.6472$ as diverse (that is, $\delta_{\text{topic}}(r_1, r_2)=1$).

For co-authorship-based diversity, we use listed publications and co-authors on reviewers' Google Scholar profiles. We calculate the authorship distance between a pair of reviewers and take reviewers with co-authorship distance $\leq$$2$ (that is, reviewers are co-authors or have a common co-author) as non-diverse (that is, $\delta_{\text{coauthor}}(r_1, r_2)=-1$), and diverse otherwise (that is, $\delta_{\text{coauthor}}(r_1, r_2)=1$). This is the same as asserting if the co-authorship vectors $\mathbf{d}(r_1)$ and $\mathbf{d}(r_2)$ have $1$'s in the same position (granted for $\mathbf{d}(r)$, you put a $1$ at reviewers' own index). 

As noted in the \autoref{subsec:data}, some profile features may be missing for some reviewers. In such cases, we treat their respective profile vectors as $0$. Subsequently, we consider their diversity term $\delta_d(r_1, r_2)$ as $0$ as well. This approach ensures a consistent estimation of effects from $-1$ for a pair of non-diverse reviewers to $1$ for a pair of diverse reviewers. 

\subsection{Effect estimation approach}
\label{subsec:approach}
To recap, for a pair of reviewers $r_1$ and $r_2$ assigned to a submitted paper $S$, we are interested in measuring the effect of diversity between the pair of reviewers ($\delta_d(r_1, r_2)$) on the redundancy or coverage of the reviews written by the two reviewers ($y(r_1, r_2; S)$), controlling for the confounders C1-C3, that is, the submitted paper ($S$), reviewers' profile ($\mathbf{d}(r_1)$ and $\mathbf{d}(r_2)$), and the reviewers' expertise ($E(r_1, S)$ and $E(r_2, S)$). 
We use the same notation ($y(r_1, r_2; S)$) for review coverage and review redundancy because the analysis procedure is the same for both quantities, so we can simply substitute $y$ for a measure of redundancy or coverage. 

We now discuss the two approaches---parametric approach and non-parametric approach that we take in our analyses.

\paragraph{\textbf{Parametric approach}} Following previous observational studies in the peer-review setting \citep{tomkins2017doubleblind, teplitskiy2019experts, stelmakh2023citeseeing}, in our parametric approach, we assume a linear model of review coverage and redundancy. 
In this linear model, we take the review coverage or redundancy as the dependent variable and the treatments (diversity along various dimensions) and confounding factors (submission, reviewers' profile, and reviewers' expertise) as the independent variables. 
\begin{align}\label{eq:linearreg}
    y(r_1, r_2; S) \sim \sum_{d\in D}\big\{\gamma_d \delta_d(r_1, r_2)\big\} + \sum_{d\in D} \big\{\beta_{d} \mathbf{d}(r_1) + \beta_{d} \mathbf{d}(r_2)\big\} + \eta_1 E(r_1, S) + \eta_2 E(r_2, S)+ \omega S + \epsilon,
\end{align}
where $\gamma_d$, $\beta_{d, 1}$, $\beta_{d, 2}$, $\eta_1$, $\eta_2$, $\omega$, and $\epsilon$ are regression parameters that need to be estimated from data. $\epsilon$ denotes the error term, which is independent of the observations. Under this assumption, $\gamma_d$ denotes the coefficient for the diversity term $\delta_d(r_1, r_2)$ for diversity dimension $d\in D$.

The main challenge in estimating the effect in \autoref{eq:linearreg} is that fully controlling for the confounder C1, that is, the submission $S$, is difficult due to the high-dimensionality of submission. The key insight to doing this analysis is that in most cases we have not two, but three or more reviews per submission. Thus, for each submitted paper, we have more than one data point. We restrict our analysis to papers that have one pair of reviewers that are diverse and one pair of reviewers that are non-diverse. We can then conduct a causal analysis across such pairs of reviewers within the submitted paper, removing the effect of $S$ in \autoref{eq:linearreg}. Essentially, without loss of generality, assume that among the three reviewers $r_1, r_2,$ and $r_3$ assigned to a submitted paper $S$, reviewers $(r_1, r_2)$ are diverse along a dimension $d^*$ and the pair $(r_1, r_3)$  are non-diverse along the dimension $d^*$, that is, $\delta_{d^*}(r_1, r_2) = 1$ and $\delta_{d^*}(r_1, r_3) = -1$. Subsequently, we can expand out \autoref{eq:linearreg} as

\begin{subequations}
\begin{equation}
    y(r_1, r_2; S) \sim \gamma_{d^*} + \sum_{\substack{d\in D \\ d\neq d^*}} \gamma_d \delta_d(r_1, r_2) + \sum_{d\in D} \beta_{d} \mathbf{d}(r_1) + \sum_{d\in D} \beta_{d} \mathbf{d}(r_2) + \eta_1 E(r_1, S) + \eta_2 E(r_2, S) + \omega S + \epsilon,
    \label{subeq:diverse}
\end{equation}
\begin{equation}
    y(r_1, r_3; S) \sim -\gamma_{d^*} + \sum_{\substack{d\in D \\ d\neq d^*}} \gamma_d \delta_d(r_1, r_3) + \sum_{d\in D} \beta_{d} \mathbf{d}(r_1) + \sum_{d\in D} \beta_{d} \mathbf{d}(r_3) + \eta_1 E(r_1, S) + \eta_2 E(r_3, S) + \omega S + \epsilon.
    \label{subeq:nondiverse}
\end{equation}
\end{subequations}
Note that in estimating the effect of diversity along dimension $d^*$, we control the interaction between the pair of reviewers along the other dimensions $d\in D$. This accounts for the confounding introduced by other dimensions of diversity. For instance, say geographical and organizational diversity are correlated, since reviewers from different geographical regions would likely have different affiliations. Subsequently, if we do not control for the differences in redundancy/coverage measures arising from geographical differences in a pair of reviewers, we may end up inflating the effect coming solely from organizational diversity. Thus, in measuring the effect of any target dimension $d^*$, we control for diversity between a pair of reviewers along all the other dimensions, along with the individual reviewers' profile features along all the dimensions (including $d^*$). 
Taking a difference between the two (that is, \autoref{subeq:diverse} and \ref{subeq:nondiverse}), we get
\begin{align}\label{eq:linearregressiondifference}
    y(r_1, r_2; S) - y(r_1, r3; S) \sim \gamma_{d^*} + \sum_{\substack{d\in D \\ d\neq d^*}} \gamma_d [\delta_d(r_1, r_2) - \delta_d(r_1, r_3)] + \sum_{d\in D} \beta_{d} [\mathbf{d}(r_2) - \mathbf{d}(r_3)] + \eta_2 [E(r_2, S) - E(r_3, S)].
\end{align}
By taking this difference, we can exclude the effect of the submission on the review redundancy and coverage measure in the linear model.
Note that the profile features and expertise of the anchor reviewer $r_1$, who is common to both the diverse pair $(r_1, r_2)$ and the non-diverse pair $(r_1, r_3)$ do not contribute to the difference in coverage and redundancy scores between the two pairs, leaving only the terms for $r_2$ and $r_3$. 
Essentially, we obtain a new linear model where the difference in redundancy and coverage measure between the diverse and non-diverse pair of reviewers serves as the dependent variable and the difference in the interaction terms between the two pairs of reviewers, along with the difference in the profile vectors and expertise of reviewers $r_2$ and $r_3$ serve as dependent variables. Subsequently, we can fit this linear model to estimate $\gamma_{d^*}$ as the effect of diversity along the dimension $d^*$ on the difference in redundancy or coverage between diverse and non-diverse pairs of reviewers. 

We fit a separate weighted linear regression model (using \texttt{python} \texttt{statsmodels} package) for each combination of diversity dimension $d^*$ and outcome measure $y$. In turn, we test for the significance of $\gamma_{d^*}$ to test for our hypotheses. 
Based on this linear model, a positive $\gamma_{d^*}$ would indicate that the respective measure of coverage or redundancy is higher for diverse reviewers along dimension $d^*\in D$ and the converse is true for a negative $\gamma_{d^*}$. 
Thus, we formally translate our hypotheses as:
\begin{itemize}
    \item \textit{Hypothesis 1:} 
    Diverse reviewer slates have \textbf{higher coverage} than non-diverse reviewer slates $\equiv$ \textbf{$\gamma_{d^*}$ is positive} when $y$ corresponds to coverage measures. 
    \item \textit{Hypothesis 2:} 
    Diverse reviewer slates have \textbf{lower redundancy} than non-diverse reviewer slates $\equiv$ \textbf{$\gamma_{d^*}$ is negative} when $y$ corresponds to redundancy measures. 
\end{itemize}

\paragraph{\textbf{Non-parametric approach}} The linearity assumption in the parametric approach, although a standard practice in observational studies, can be a strong modeling assumption. Thus, we supplement our parametric analysis with a non-parametric approach. In this approach, we match two reviewer slates---one diverse and one non-diverse---across all covariates and assess the difference in the review coverage and redundancy between these groups. However, since we consider multiple axes of diversity and reviewer profiles, it is difficult to find a sufficient number of submissions that have both a diverse pair of reviewers and a non-diverse pair of reviewers, with all the other features exactly matching between the two. To circumvent this, we instead employ propensity score matching \citep{rosenbaum1983propensitymatching}. 

In propensity score matching, the goal is to match a diverse reviewer slate with a non-diverse reviewer slate such that their propensity score (the probability of the treatment given all other covariates) is similar. This matching helps remove potential bias in treatment assignments and assess differences in review coverage and redundancy between diverse and non-diverse slates. As with the parametric approach, we control for the effect of submission by performing the propensity matching within a submitted paper ($S$). We first train a logistic regression model (using \texttt{python} \texttt{statsmodels} package) to predict the treatment given the other covariates (excluding $S$). For each submitted paper, we then match each diverse reviewer slate with a non-diverse reviewer slate if their propensity scores are similar (within a $0.1$ range). We discard cases where none of the reviewer pairs within a paper meets this criterion. This results in $36$, $174$, $113$, $106$, and $96$ matched data points for organization, geographical location, seniority, topic, and co-authorship diversity, respectively. We exclude organizational diversity from this analysis due to the low number of matched data points.

Finally, to assess the differences in review coverage and redundancy between the diverse and non-diverse slates, we calculate the difference in the mean of review coverage or redundancy between the two matched groups. We measure statistical significance using a permutation test \citep{fisher1936permutation}. The details of the non-parametric approach can be found in \autoref{appendix:non-parametric}.

\section{Results}\label{sec:results}

\autoref{tab:effect} presents the effect size $\gamma_{d^*}$ of reviewer diversity along dimensions on different measures of review coverage and redundancy with our parametric approach. Our non-parametric analysis, when available, corroborates these findings.\footnote{We excluded organizational diversity from non-parametric analysis due to a low number of matched data points. }\footnote{The p-values for the parametric and the non-parametric approach are not exactly the same, but the findings align---whichever effect is significant in the linear analysis is also significant in the non-parametric analysis, and vice versa. }
We do not include them here for brevity. All results are reported at a significance threshold of $0.01$ obtained after multiple testing corrections using the Benjamini-Hochberg Procedure with a false discovery rate of $0.05$ \citep{fdr1995benjamini}.
Our analysis reveals that different measures of diversity have varying effects on review coverage and redundancy.

\begin{table*}[t]
\scriptsize
    \centering
    \begin{tabular}{lr|>{\columncolor{lightpink}}c>{\columncolor{lightpink}}c|>{\columncolor{lightpink}}c>{\columncolor{lightpink}}c|>{\columncolor{lightpurple}}c>{\columncolor{lightpurple}}c|>{\columncolor{lightpurple}}c>{\columncolor{lightpurple}}c}
    \toprule
        & & \multicolumn{4}{>{\columncolor{lightpink}}c|}{Coverage} & \multicolumn{4}{>{\columncolor{lightpurple}}c}{Redundancy} \\
        \midrule
        & & \multicolumn{2}{>{\columncolor{lightpink}}c|}{Type Coverage} & \multicolumn{2}{>{\columncolor{lightpink}}c|}{Paper Coverage} & & & \multicolumn{2}{>{\columncolor{lightpurple}}c}{Weighted Semantic}
        \\
        Diversity ($\downarrow$) & \# & Argument & Aspect & Lexical & Semantic & Lexical & Semantic & Argument & Aspect \\
        \midrule
        Organization & $237$ & \hphantom{$-$}$0.0039$\hphantom{$^{*}$} & \hphantom{$-$}$0.0037$\hphantom{$^{*}$} & \hphantom{$-$}$0.0006$\hphantom{$^{*}$} & \hphantom{$-$}$0.0036$\hphantom{$^{*}$} & $-0.0218^{*}$ & $-0.0078^{*}$ & $-0.0024$\hphantom{$^{*}$} & $-0.0014$\hphantom{$^{*}$} \\
        Geographical & $1432$ & \hphantom{$-$}$0.0022$\hphantom{$^{*}$} & $-0.0004$\hphantom{$^{*}$} & \hphantom{$-$}$0.0007$\hphantom{$^{*}$} & $-0.0006$\hphantom{$^{*}$} & $-0.0032$\hphantom{$^{*}$} & $-0.0014$\hphantom{$^{*}$} & \hphantom{$-$}$0.0000$\hphantom{$^{*}$} & \hphantom{$-$}$0.0002$\hphantom{$^{*}$} \\
        Seniority & $1127$ & \hphantom{$-$}$0.0058^{*}$ & \hphantom{$-$}$0.0074^{*}$ & \hphantom{$-$}$0.0004$\hphantom{$^{*}$} & $-0.0016$\hphantom{$^{*}$} & $-0.0061^{*}$ & $-0.0015^{*}$ & $-0.0001$\hphantom{$^{*}$} & $-0.0002$\hphantom{$^{*}$} \\
        Topical & $387$ & $-0.1307$\hphantom{$^{*}$} & $-0.1028$\hphantom{$^{*}$} & \hphantom{$-$}$0.1098^{*}$ & \hphantom{$-$}$0.0929^{*}$ & $-0.0524^{*}$ & $-0.0290^{*}$ & \hphantom{$-$}$0.0068$\hphantom{$^{*}$} & \hphantom{$-$}$0.0085$\hphantom{$^{*}$} \\
        Co-authorship & $314$ & \hphantom{$-$}$0.0064^{*}$ & \hphantom{$-$}$0.0059^{*}$ & \hphantom{$-$}$0.0084$ \hphantom{$^{*}$} & \hphantom{$-$}$0.0057$\hphantom{$^{*}$} & $-0.0258^{*}$ & $-0.0097^{*}$ & $-0.0018^{*}$ & $-0.0012^{*}$ \\
        \bottomrule
    \end{tabular}
    \caption{Effect of reviewer diversity along different dimensions on review redundancy and coverage ($\gamma_{d^*}$ in Eq. \protect\ref{eq:linearregressiondifference}) using parametric approach. We perform Benjamini-Hochberg correction to counteract the multiple comparison problem with a false discovery threshold of $0.05$ ($^*$ denotes significance at the p-value threshold of $0.01$ obtained after multiple testing correction). For hypothesis 1---diverse slates have higher review coverage---we expect a positive $\gamma_{d^*}$ for review coverage. For hypothesis 2---diverse slates have lower review redundancy---we expect a negative $\gamma_{d^*}$ for review redundancy.
    }
    \label{tab:effect}
\end{table*}

\subsection{Hypothesis 1: Diverse reviewers have higher review coverage.}

\paragraph{\textit{Co-authorship and seniority-based diversity in reviewer slate yield reviews with higher type coverage.}}
Our analysis reveals a positive and significant effect of co-authorship and seniority-based diversity on the type coverage of reviews. Specifically, co-authorship diversity exhibits an effect of $0.0064$ (p-value=$0.009$) for argument-type coverage and $0.0059$ (p-value=$0.008$) for aspect-type coverage. 
Similarly, seniority-based diversity exhibits an effect of $0.0058$ (p-value=$0.003$) for argument-type coverage and $0.0074$ (p-value=$0.007$) for aspect-type coverage. 
This suggests that reviewers from diverse publication networks or seniority levels tend to provide reviews that evaluate a submitted paper along varying axes. This aligns with the intuition that reviewers from different publication networks and seniority-level may differ in their evaluation criteria, leading to a focus on different aspects of a review. Notably, organizational, geographical, and topical diversity, do not exhibit any significant effect on the two measures of type coverage. 

\paragraph{\textit{Topical diversity in reviewer slates yield reviews with higher paper coverage.}}
Our analysis reveals a positive and significant effect of topical diversity on the paper's lexical and semantic coverage. Specifically, topical diversity exhibits an effect of $0.1098$ (p-value=$0.004$) on lexical paper coverage and $0.0929$ (p-value=$0.01$) on semantic paper coverage. This suggests that reviews of topically diverse reviewers cover more elements of a paper, as gleaned through the paper's abstract. This aligns with the intuition, as topically diverse reviewers may possess diverse expertise or interests, leading to a focus on different aspects of a paper. Other axes of diversity, namely, organization, geographical, co-authorship, and seniority-based diversity, do not exhibit any significant effect on the two measures of paper coverage.

\paragraph{\textit{Why are findings different for paper and type coverage measures?}}
It's important to note that although both type and paper coverage address coverage in reviews, they serve distinct purposes. Type coverage evaluates how many dimensions of a review are addressed by a pair of reviews, while paper coverage assesses how many dimensions of the paper are addressed. Consequently, type coverage considers aspects of reviews like evaluating the motivation, originality, soundness, or applicability of a submitted paper. On the other hand, paper coverage examines coverage of different elements of the paper outlined in its abstract. For example, an abstract might highlight the paper's primary contribution as developing a new evaluation paradigm, discussing its validity, and presenting empirical evidence of its effectiveness. Different reviews may prioritize assessing the validity of the evaluation paradigm versus the validity of the empirical experiments. In this case, type coverage would only count soundness as the dimension of review, but paper coverage would reveal the different elements of paper covered across the pair of reviews. Thus, type and paper coverage highlight complementary properties of reviews. 

In conclusion, high paper coverage for topically diverse reviewers reflects their broader expertise or interests, leading to a focus on different parts of the paper. Conversely, high type coverage for co-authorship and seniority-based diverse reviewers reflects their varied evaluation criteria, resulting in coverage of diverse aspects of a review.

\subsection{Hypothesis 2: Diverse reviewers have lower review redundancy.} 

\paragraph{\textit{Organizational, co-authorship, topical, and seniority-based diversity in reviewer slates yield reviews with low redundancy.}}
Our analysis reveals that organizational, co-authorship, topical, and seniority-based diversity exhibit a negative and significant effect on lexical and semantic redundancy. Specifically, organization diversity exhibits an effect of $-0.0218$ (p-value=$0.000$) on lexical redundancy and $-0.0078$ (p-value=$0.003$) on semantic redundancy measures. Co-authorship-based diversity exhibits an effect of $-0.0258$ (p-value=$0.000$) on lexical redundancy and $-0.0097$ (p-value=$0.004$) on semantic redundancy measures. Topical diversity exhibits an effect of $-0.0524$ (p-value=$0.01$) on lexical redundancy and $-0.0290$ (p-value=$0.009$) on semantic redundancy. Lastly, seniority-based diversity exhibits an effect of $-0.0061$ (p-value=$0.006$) on lexical redundancy and $-0.0015$ (p-value=$0.007$) on semantic redundancy measures. This indicates that reviewers from diverse organizations, publication networks, topics, and seniority levels have lesser lexical and semantic redundancy in their reviews. 

\paragraph{\textit{Co-authorship diversity in reviewer slate alone yield low redundancy weighted by review types.}}
Considering the weighted semantic redundancy, we find that co-authorship-based diversity has a negative and significant effect on review redundancy. Specifically, co-authorship-based diversity exhibits an effect of $-0.0018$ (p-value=$0.004$) for argument-wise weighted semantic redundancy and $-0.0012$ (p-value=$0.006$) for aspect-wise weighted semantic redundancy. This indicates that reviewers from diverse publication networks have lesser redundancy in reviews, even within an aspect or argument type. Other dimensions of diversity do not exhibit any significant effect on redundancy within a review type. 

\paragraph{\textit{Why are findings different for weighted and unweighted semantic redundancy measures?}}
The unweighted semantic redundancy considers the overall overlap in reviews. Thus, it simply measures how similar the reviews are semantically. 
As hinted at previously, lower redundancy may naturally accompany high coverage since broader coverage in reviews is likely, but not necessarily, a result of different reviews focusing on different review criteria, thus resulting in less overlap (that is, low redundancy). This is confirmed by our results as, unsurprisingly, the three axes of diversity with larger coverage along some measure of coverage (co-authorship, topical, and seniority-based diversity) have lower lexical and semantic redundancy.  Weighted semantic redundancy however specifically assesses redundancy in reviews weighted by the similarity in their argument- and aspect-types. In this case, higher similarity between reviews that focus on the same review criterion results in higher weighted semantic redundancy. Thus, low weighted semantic redundancy indicates that reviews from diverse pairs of reviewers offer different perspectives even within the same review criterion.

In conclusion, a low redundancy for organizational, co-authorship, topical, and seniority-based diversity reflects generally a higher variability in the reviews. Going a step further, a lower weighted redundancy even within different aspects and argument types for co-authorship-based diversity indicates that reviews written by reviewers belonging to different publication networks not only have high type coverage but also have low redundancy within reviews of specific types.

\subsection{Other findings}

\paragraph{\textit{Does high coverage and low redundancy in reviews of diverse reviewer slates a result of potentially low quality reviews?}}
One possible objection to this analysis is that high coverage and low redundancy in diverse reviewer slates could stem from one of the reviews being of low quality. For instance, topically diverse reviewers might have a low redundancy because one of their reviews has nothing to do with the paper. While we aim for non-redundant and high coverage reviews, peer-reviewing should still result in useful reviews. To rule out this alternate hypothesis, that is, reviews from diverse slates are worse, we measure Pearson's correlation of diversity with an alternate quality indicator---meta-reviewers' rating of the reviews (on a scale of $1$--$5$). The Pearson's r correlation coefficient \citep{pearson1985note} between the slate diversity and review rating is not significant for any measure of diversity: Organization ($-0.065$, p-value=$0.2119$), Geographical ($-0.016$, p-value=$0.4423$), Co-authorship ($0.006$, p-value=$0.8893$), Topical ($0.016$, p-value=$0.6645$), and Seniority ($0.000$, p-value=$0.9821$). 

\paragraph{\textit{Reviews from more expert reviewers slates yield higher coverage and lower redundancy.}}
One aspect of our analysis is controlling for the reviewers' expertise with respect to the submitted paper when estimating the effect of reviewer diversity on review coverage and redundancy. In addition to studying how the diversity of reviewer slates (our treatment) affects the review coverage and redundancy (our outcomes), we also assess the effect of the reviewer expertise (a confounder) on the outcomes.
We observe that reviewer expertise has a positive effect on review coverage and a negative effect on review redundancy for all measures of coverage and redundancy. This suggests that when one or both reviewers in the pair possess greater expertise with respect to the submission, there is an overall increase in both type and paper coverage, coupled with a decrease in lexical, semantic, and weighted semantic redundancy. This validates the common intuition that reviewers with more expertise would yield better, more valuable reviews. In addition to this factor, our paper proposes other group-based considerations in the reviewer assignment process.

\section{Discussion}
Our analysis presents mixed findings for the two hypotheses investigated in our study. Hypothesis 1, suggesting that diverse reviewer slates result in higher review coverage, is supported by co-authorship and seniority based diversity measures for type coverage, as well as by topical diversity measures for paper coverage. Hypothesis 2, suggesting that diverse slates have lower review redundancy, is supported by all diversity measures (organizational, co-authorship, topical, and seniority), except for geographical diversity, for lexical and semantic redundancy measures. Notably, co-authorship-based diversity demonstrates a consistent reduction in redundancy, both broadly (in overall lexical and semantic redundancy) and locally (weighted semantic redundancy within specific arguments and aspects). 

Echoing literature in teams and collaboration pertaining to group decision-making, diversity among reviewers assigned to a submitted paper has recently been proposed as an essential consideration in peer-reviewing \citep{shah2022overview}. 
Reviewers in specific research niches, publication networks, or even geographical regions have different preferences and practices, which might add bias to the evaluation. \citet{shah2022overview} argues for diversity in reviewer assignments to ensure higher coverage in evaluating interdisciplinary research. 
A popular machine learning conference, Association for the Advancement of Artificial Intelligence (AAAI 2021), with about $7,900$ submitted papers, included soft geographical and co-authorship constraints in the reviewer assignment algorithm for promoting geographical and co-authorship based diversity \citep{AAAI2024papermatching}, albeit with the goal of mitigating collusion rings, different from the review utility criteria discussed in this paper. 

Wider adoption of such strategies, however, requires a deeper understanding of the benefits they offer. Introducing numerous constraints on reviewer assignments can prove impractical and labor-intensive for manual assignments, and computationally demanding for automated systems. Our study examines the impacts of diversity on the peer-review process using data from past conferences, offering insights into potentially beneficial constraints.
Our study reveals that co-authorship-based diversity increases coverage in reviews and reduces redundancy, suggesting the potential benefits of assigning reviewers from diverse publication networks to a paper. Additionally, topical and seniority-based diversity may also contribute to increased coverage and reduced redundancy for certain measures of coverage and redundancy. Lastly, organization-based diversity yields improvement in only some measures of redundancy and does not significantly enhance coverage. Importantly, despite some previous adoptions of geographical diversity in reviewer assignments (e.g., AAAI 2021), we do not find any impact of geographical diversity on review redundancy and coverage. We hope that our findings will provide valuable insights into the efficacy of different axes of reviewer diversity, potentially guiding their adoption in reviewer assignments. 

\paragraph{Policy Recommendations} \textit{Based on our findings detailed above, we present the following actionable considerations for reviewer assignment in conference peer-review:
}

\begin{itemize}
\item {In addition to optimizing individual reviewer expertise in relation to a given submission, choosing diverse \textit{slates} of reviewers can lead to better reviews with higher coverage and lower redundancy.}
\item To ensure more comprehensive peer reviews, it is recommended that the slate of reviewers assigned to a paper be composed of individuals who are topically diverse, have different seniority levels, and are from distinct publication networks (that is, they are not co-authors and do not share common co-authors). Our study finds that such reviewer slates provide a collection of reviews with broader coverage of both the paper and the review criteria.
\item To ensure wider perspectives in reviews, it is recommended that the slate of reviewers assigned to a paper be composed of reviewers from distinct publication networks. 
Our study finds that reviewers from diverse publication networks provide reviews with varying perspectives, even within specific review criteria, potentially offering more insights for the authors to improve the submission and for editors to evaluate it more comprehensively.
\end{itemize}
{More generally, the analyses we conduct can also be extended to other outcomes of interest, potentially informing key policies for the reviewer-assignment process.}

\paragraph{Limitations} While we advocate for higher coverage and lower redundancy in reviews across a slate of reviewers, we acknowledge that there are associated challenges. On one hand, higher coverage can help ensure that reviewers broadly evaluate the merits and demerits of the work \citep{porter1985peerinterdisciplinary, Laudel2006Interdisciplinary, shah2022overview}. On the other hand, ``the broader the intellectual territory covered, the less consensus there will be on the ranking'' \citep{brooks1978ResearchPriorities}. Similarly, while higher redundancy may indicate bias in reviews, it can simplify the decision-making process for the editorial board of the venue. Despite these challenges, considering that many venues integrate textual comments and discussions into their peer-review process, a broader perspective in the reviews could be beneficial overall. The hope is that detailed feedback and discussion, originating from more diverse and less biased sources, can enable editors or area chairs to formulate a more comprehensive and expansive evaluation of the submitted papers.

One limitation of our work is that we focus solely on certain aspects of review utility, namely redundancy and coverage, operationalized through specific measures. We recognize that there may be other factors that are crucial, either generally or specifically to a particular conference or journal, which we haven't addressed. While our paper discusses the potential benefits of specific dimensions of reviewer diversity, the primary contribution of our paper lies in demonstrating systematic analysis using observational data derived from existing peer-review datasets. Such systematic analyses can inform adjustments to the reviewer assignment process or serve as a foundation for conducting controlled experiments aimed at improving the peer-review system. 

Another limitation of our work is that it focuses on the initial reviews, and does not analyze the group dynamics that come into play when reviewers are allowed to interact. While the initial reviews themselves play a pivotal role in the review process, many peer-review systems in paper and proposal review also involve a subsequent discussion among reviewers. There are several studies on various aspects of peer-reviewer discussions~\citep{obrecht2007examining, fogelholm2012panel, pier2017your, stelmakh2020herding, rastogi2024randomized}, and analyzing the effects of diversity can shed more light on improving the efficacy of the discussion processes.

\section*{Acknowledgments}

We would like to thank the current and former members of the UMD CLIP lab, especially Alexander Hoyle and Connor Baumler for their useful suggestions and feedback. This research was supported by NSF 1942124 and ONR N000142212181.

\bibliographystyle{acl_natbib}
\bibliography{references}

\begin{thebibliography}{50}
\expandafter\ifx\csname natexlab\endcsname\relax\def\natexlab#1{#1}\fi

\bibitem[{Abma-Schouten et~al.(2023)Abma-Schouten, Gijbels, Reijmerink, and
  Meijer}]{schouten2023broaderpanels}
Rebecca Abma-Schouten, Joey Gijbels, Wendy Reijmerink, and Ingeborg Meijer.
  2023.
\newblock \href {https://doi.org/10.1093/scipol/scad009} {{Evaluation of
  research proposals by peer review panels: broader panels for broader
  assessments?}}
\newblock \emph{Science and Public Policy}, 50(4):619--632.

\bibitem[{Bantel and Jackson(1989)}]{bantel1989top}
Karen~A Bantel and Susan~E Jackson. 1989.
\newblock \href
  {https://onlinelibrary.wiley.com/doi/abs/10.1002/smj.4250100709} {Top
  management and innovations in banking: Does the composition of the top team
  make a difference?}
\newblock \emph{Strategic management journal}, 10(S1):107--124.

\bibitem[{Benjamini and Hochberg(1995)}]{fdr1995benjamini}
Yoav Benjamini and Yosef Hochberg. 1995.
\newblock \href {http://www.jstor.org/stable/2346101} {Controlling the false
  discovery rate: A practical and powerful approach to multiple testing}.
\newblock \emph{Journal of the Royal Statistical Society. Series B
  (Methodological)}, 57(1):289--300.

\bibitem[{Bianchi and Squazzoni(2015)}]{bianchi2015three}
Federico Bianchi and Flaminio Squazzoni. 2015.
\newblock \href {https://dl.acm.org/doi/10.5555/2888619.2889159} {Is three
  better than one? {S}imulating the effect of reviewer selection and behavior
  on the quality and efficiency of peer review}.
\newblock In \emph{Proceedings of the 2015 Winter Simulation Conference}, WSC
  '15, page 4081–4089. IEEE Press.

\bibitem[{Blei et~al.(2003)Blei, Ng, and Jordan}]{blei2001lda}
David Blei, Andrew Ng, and Michael Jordan. 2003.
\newblock \href {https://dl.acm.org/doi/abs/10.5555/944919.944937} {Latent
  dirichlet allocation}.
\newblock \emph{The Journal of Machine Learning Research}, 3(1):993–1022.

\bibitem[{Brooks(1978)}]{brooks1978ResearchPriorities}
Harvey Brooks. 1978.
\newblock \href {http://www.jstor.org/stable/20024552} {The problem of research
  priorities}.
\newblock \emph{Daedalus}, 107(2):171--190.

\bibitem[{Charlin and Zemel(2013)}]{charlin2013toronto}
Laurent Charlin and Richard Zemel. 2013.
\newblock \href {https://openreview.net/forum?id=caynafZAnBafx} {The toronto
  paper matching system: {A}n automated paper-reviewer assignment system}.
\newblock In \emph{Proceedings of the International Conference on Machine
  Learning Workshop on Peer Reviewing and Publishing Models}.

\bibitem[{Chubin and Hackett(1990)}]{chubin1990peerless}
Daryl~E Chubin and Edward~J Hackett. 1990.
\newblock \emph{Peerless science: Peer review and US science policy}.
\newblock State University of New York Press.

\bibitem[{Fisher(1936)}]{fisher1936permutation}
Ronald~Aylmer Fisher. 1936.
\newblock \href {https://pmc.ncbi.nlm.nih.gov/articles/PMC2458144/} {Design of
  experiments}.
\newblock \emph{British Medical Journal}, 1(3923):554.

\bibitem[{Fogelholm et~al.(2012)Fogelholm, Leppinen, Auvinen, Raitanen,
  Nuutinen, and V{\"a}{\"a}n{\"a}nen}]{fogelholm2012panel}
Mikael Fogelholm, Saara Leppinen, Anssi Auvinen, Jani Raitanen, Anu Nuutinen,
  and Kalervo V{\"a}{\"a}n{\"a}nen. 2012.
\newblock \href
  {https://www.sciencedirect.com/science/article/pii/S089543561100148X} {Panel
  discussion does not improve reliability of peer review for medical research
  grant proposals}.
\newblock \emph{Journal of clinical epidemiology}, 65(1):47--52.

\bibitem[{Grimaldo and Paolucci(2013)}]{grimaldo2013simulation}
Francisco Grimaldo and Mario Paolucci. 2013.
\newblock \href
  {https://www.worldscientific.com/doi/abs/10.1142/S0219525913500045} {A
  simulation of disagreement for control of rational cheating in peer review}.
\newblock \emph{Advances in Complex Systems}, 16(07):1350004.

\bibitem[{Hargens and Herting(1990)}]{hargens1990neglected}
Lowell Hargens and Jerald Herting. 1990.
\newblock \href {https://doi.org/10.1007/BF02130467} {Neglected considerations
  in the analysis of agreement among journal referees}.
\newblock \emph{Scientometrics}, 19(1-2):91--106.

\bibitem[{Hoffman et~al.(1962)Hoffman, Harburg, and
  Maier}]{hoffmann1962differences}
L~Richard Hoffman, Ernest Harburg, and Norman~RF Maier. 1962.
\newblock \href {https://doi.org/10.1037/h0045952} {Differences and
  disagreement as factors in creative group problem solving.}
\newblock \emph{The Journal of Abnormal and Social Psychology}, 64(3):206.

\bibitem[{Hofstra et~al.(2020)Hofstra, Kulkarni, Galvez, He, Jurafsky, and
  McFarland}]{Hofstra2020diversityparadox}
Bas Hofstra, Vivek~V. Kulkarni, Sebastian Munoz-Najar Galvez, Bryan He, Dan
  Jurafsky, and Daniel~A. McFarland. 2020.
\newblock \href {https://doi.org/10.1073/pnas.1915378117} {The
  diversity–innovation paradox in science}.
\newblock \emph{Proceedings of the National Academy of Sciences},
  117(17):9284--9291.

\bibitem[{Hong and Page(2004)}]{hong2004problemsolving}
Lu~Hong and Scott~E. Page. 2004.
\newblock \href {https://doi.org/10.1073/pnas.0403723101} {Groups of diverse
  problem solvers can outperform groups of high-ability problem solvers}.
\newblock \emph{Proceedings of the National Academy of Sciences},
  101(46):16385--16389.

\bibitem[{Hua et~al.(2019)Hua, Nikolov, Badugu, and Wang}]{hua2019argument}
Xinyu Hua, Mitko Nikolov, Nikhil Badugu, and Lu~Wang. 2019.
\newblock \href {"https://aclanthology.org/N19-1219"} {Argument mining for
  understanding peer reviews}.
\newblock In \emph{Proceedings of the 2019 Conference of the North American
  Chapter of the Association for Computational Linguistics: Human Language
  Technologies, Volume 1}, pages 2131--2137.

\bibitem[{Jackson et~al.(1995)Jackson, May, Whitney, Guzzo, and
  Salas}]{jackson1995understanding}
Susan~E Jackson, Karen~E May, Kristina Whitney, Richard~A Guzzo, and Eduardo
  Salas. 1995.
\newblock Understanding the dynamics of diversity in decision-making teams.
\newblock \emph{Team effectiveness and decision making in organizations},
  204:261.

\bibitem[{Jecmen et~al.(2020)Jecmen, Zhang, Liu, Shah, Conitzer, and
  Fang}]{jecmen2020mitigating}
Steven Jecmen, Hanrui Zhang, Ryan Liu, Nihar Shah, Vincent Conitzer, and Fei
  Fang. 2020.
\newblock \href
  {https://proceedings.neurips.cc/paper_files/paper/2020/file/93fb39474c51b8a82a68413e2a5ae17a-Paper.pdf}
  {Mitigating manipulation in peer review via randomized reviewer assignments}.
\newblock In \emph{Advances in Neural Information Processing Systems},
  volume~33, pages 12533--12545. Curran Associates, Inc.

\bibitem[{Jefferson et~al.(2002{\natexlab{a}})Jefferson, Alderson, Wager, and
  Davidoff}]{jefferson2002effects}
Tom Jefferson, Philip Alderson, Elizabeth Wager, and Frank Davidoff.
  2002{\natexlab{a}}.
\newblock \href {https://doi.org/10.1001/jama.287.21.2784} {Effects of
  editorial peer review: a systematic review}.
\newblock \emph{JAMA}, 287(21):2784--2786.

\bibitem[{Jefferson et~al.(2002{\natexlab{b}})Jefferson, Wager, and
  Davidoff}]{jefferson2002quality}
Tom Jefferson, Elizabeth Wager, and Frank Davidoff. 2002{\natexlab{b}}.
\newblock \href {https://doi.org/10.1001/jama.287.21.2786} {{Measuring the
  Quality of Editorial Peer Review}}.
\newblock \emph{JAMA}, 287(21):2786--2790.

\bibitem[{Kuznetsov et~al.(2024)Kuznetsov, Afzal, Dercksen, Dycke, Goldberg,
  Hope, Hovy, Kummerfeld, Lauscher, Leyton-Brown
  et~al.}]{kuznetsov2024whatcannlpdo}
Ilia Kuznetsov, Osama~Mohammed Afzal, Koen Dercksen, Nils Dycke, Alexander
  Goldberg, Tom Hope, Dirk Hovy, Jonathan~K Kummerfeld, Anne Lauscher, Kevin
  Leyton-Brown, et~al. 2024.
\newblock \href {http://arxiv.org/abs/2405.06563} {What can natural language
  processing do for peer review?}
\newblock \emph{arXiv preprint arXiv:2405.06563}.

\bibitem[{Laudel(2006)}]{Laudel2006Interdisciplinary}
Grit Laudel. 2006.
\newblock \href {https://doi.org/10.3152/147154406781776048} {{Conclave in the
  Tower of Babel: {H}ow peers review interdisciplinary research proposals}}.
\newblock \emph{Research Evaluation}, 15(1):57--68.

\bibitem[{Lee(2012)}]{Lee2012Kuhnian}
Carole~J. Lee. 2012.
\newblock \href {https://doi.org/10.1086/667841} {A kuhnian critique of
  psychometric research on peer review}.
\newblock \emph{Philosophy of Science}, 79(5):859–870.

\bibitem[{Lee et~al.(2013)Lee, Sugimoto, Zhang, and Cronin}]{lee2013bias}
Carole~J Lee, Cassidy~R Sugimoto, Guo Zhang, and Blaise Cronin. 2013.
\newblock \href {https://doi.org/10.1002/asi.22784} {Bias in peer review}.
\newblock \emph{Journal of the American Society for Information Science and
  Technology}, 64(1):2--17.

\bibitem[{Levine et~al.(2014)Levine, Apfelbaum, Bernard, Bartelt, Zajac, and
  Stark}]{levine2014ethnic}
Sheen~S Levine, Evan~P Apfelbaum, Mark Bernard, Valerie~L Bartelt, Edward~J
  Zajac, and David Stark. 2014.
\newblock \href {https://doi.org/10.1073/pnas.1407301111} {Ethnic diversity
  deflates price bubbles}.
\newblock \emph{Proceedings of the National Academy of Sciences},
  111(52):18524--18529.

\bibitem[{Leyton-Brown et~al.(2024)Leyton-Brown, Mausam, Nandwani, Zarkoob,
  Cameron, Newman, and Raghu}]{AAAI2024papermatching}
Kevin Leyton-Brown, Mausam, Yatin Nandwani, Hedayat Zarkoob, Chris Cameron,
  Neil Newman, and Dinesh Raghu. 2024.
\newblock \href {https://doi.org/https://doi.org/10.1016/j.artint.2024.104119}
  {Matching papers and reviewers at large conferences}.
\newblock \emph{Artificial Intelligence}, 331:104119.

\bibitem[{Murray et~al.(2018)Murray, Siler, Larivi{\`e}re, Chan, Collings,
  Raymond, and Sugimoto}]{murray2018international}
Dakota Murray, Kyle Siler, Vincent Larivi{\`e}re, Wei~Mun Chan, Andrew~M.
  Collings, Jennifer Raymond, and Cassidy~R. Sugimoto. 2018.
\newblock \href {https://doi.org/10.1101/400515} {Gender and international
  diversity improves equity in peer review}.
\newblock \emph{bioRxiv}.

\bibitem[{Obrecht et~al.(2007)Obrecht, Tibelius, and
  D'Aloisio}]{obrecht2007examining}
Michael Obrecht, Karl Tibelius, and Guy D'Aloisio. 2007.
\newblock \href {https://doi.org/10.3152/095820207X223785} {Examining the value
  added by committee discussion in the review of applications for research
  awards}.
\newblock \emph{Research Evaluation}, 16(2):79--91.

\bibitem[{Olbrecht and Bornmann(2010)}]{olbrecht2010panelpeerreview}
Meike Olbrecht and Lutz Bornmann. 2010.
\newblock \href {https://doi.org/10.3152/095820210X12809191250762} {{Panel peer
  review of grant applications: what do we know from research in social
  psychology on judgment and decision-making in groups?}}
\newblock \emph{Research Evaluation}, 19(4):293--304.

\bibitem[{Page(2008)}]{page2008perspectives}
Scott Page. 2008.
\newblock \emph{The difference: How the power of diversity creates better
  groups, firms, schools, and societies-new edition}.
\newblock Princeton University Press.

\bibitem[{{Pearson}(1895)}]{pearson1985note}
Karl {Pearson}. 1895.
\newblock \href {https://www.jstor.org/stable/115794} {{Note on Regression and
  Inheritance in the Case of Two Parents}}.
\newblock \emph{Proceedings of the Royal Society of London Series I},
  58:240--242.

\bibitem[{Pier et~al.(2017)Pier, Raclaw, Kaatz, Brauer, Carnes, Nathan, and
  Ford}]{pier2017your}
Elizabeth Pier, Joshua Raclaw, Anna Kaatz, Markus Brauer, Molly Carnes,
  Mitchell Nathan, and Cecilia Ford. 2017.
\newblock \href {https://doi.org/10.1093/reseval/rvw025} {Your comments are
  meaner than your score: {s}core calibration talk influences intra-and
  inter-panel variability during scientific grant peer review}.
\newblock \emph{Research Evaluation}, 26(1):1--14.

\bibitem[{Porter and Rossini(1985)}]{porter1985peerinterdisciplinary}
Alan~L. Porter and Frederick~A. Rossini. 1985.
\newblock \href {https://doi.org/10.1177/016224398501000304} {Peer review of
  interdisciplinary research proposals}.
\newblock \emph{Science, Technology, \& Human Values}, 10(3):33--38.

\bibitem[{Rastogi et~al.(2024)Rastogi, Song, Jin, Stelmakh, Daum{\'e}~III,
  Zhang, and Shah}]{rastogi2024randomized}
Charvi Rastogi, Xiangchen Song, Zhijing Jin, Ivan Stelmakh, Hal Daum{\'e}~III,
  Kun Zhang, and Nihar~B Shah. 2024.
\newblock \href {https://arxiv.org/abs/2403.01015} {A randomized controlled
  trial on anonymizing reviewers to each other in peer review discussions}.
\newblock \emph{arXiv preprint arXiv:2403.01015}.

\bibitem[{Reimers and Gurevych(2019)}]{reimers2019sbert}
Nils Reimers and Iryna Gurevych. 2019.
\newblock \href {https://aclanthology.org/D19-1410/} {Sentence-{BERT}: Sentence
  embeddings using {S}iamese {BERT}-networks}.
\newblock In \emph{Proceedings of the 2019 Conference on Empirical Methods in
  Natural Language Processing and the 9th International Joint Conference on
  Natural Language Processing (EMNLP-IJCNLP)}, pages 3982--3992.

\bibitem[{Reynolds and Lewis(2017)}]{reynolds2017teams}
Alison Reynolds and David Lewis. 2017.
\newblock \href
  {https://hbr.org/2017/03/teams-solve-problems-faster-when-theyre-more-cognitively-diverse}
  {Teams solve problems faster when they’re more cognitively diverse}.
\newblock \emph{Harvard Business Review}, 30:1--8.

\bibitem[{R\"{o}der et~al.(2015)R\"{o}der, Both, and
  Hinneburg}]{roder2015topiccoherence}
Michael R\"{o}der, Andreas Both, and Alexander Hinneburg. 2015.
\newblock \href {https://doi.org/10.1145/2684822.2685324} {Exploring the space
  of topic coherence measures}.
\newblock In \emph{Proceedings of the Eighth ACM International Conference on
  Web Search and Data Mining}, WSDM '15, page 399–408, New York, NY, USA.
  Association for Computing Machinery.

\bibitem[{Rosenbaum and Rubin(1983)}]{rosenbaum1983propensitymatching}
Paul~R Rosenbaum and Donald~B Rubin. 1983.
\newblock \href {https://doi.org/10.1093/biomet/70.1.41} {The central role of
  the propensity score in observational studies for causal effects}.
\newblock \emph{Biometrika}, 70(1):41--55.

\bibitem[{Sanh et~al.(2019)Sanh, Debut, Chaumond, and
  Wolf}]{sanh2019distilbert}
Victor Sanh, Lysandre Debut, Julien Chaumond, and Thomas Wolf. 2019.
\newblock \href {https://arxiv.org/abs/1910.01108} {Distil{BERT}, a distilled
  version of {BERT}: smaller, faster, cheaper and lighter}.
\newblock \emph{arXiv preprint arXiv:1910.01108}.

\bibitem[{Saveski et~al.(2023)Saveski, Jecmen, Shah, and
  Ugander}]{saveski2023counterfactual}
Martin Saveski, Steven Jecmen, Nihar Shah, and Johan Ugander. 2023.
\newblock \href
  {https://proceedings.neurips.cc/paper_files/paper/2023/file/b7d795e655c1463d7299688d489e8ef4-Paper-Conference.pdf}
  {Counterfactual evaluation of peer-review assignment policies}.
\newblock In \emph{Advances in Neural Information Processing Systems},
  volume~36, pages 58765--58786. Curran Associates, Inc.

\bibitem[{Shah(2022)}]{shah2022overview}
Nihar~B. Shah. 2022.
\newblock \href {https://doi.org/10.1145/3528086} {Challenges, experiments, and
  computational solutions in peer review}.
\newblock \emph{Commun. ACM}, 65(6):76–87.

\bibitem[{Sommers(2006)}]{sommers2006racial}
Samuel~R Sommers. 2006.
\newblock \href {https://doi.org/10.1037/0022-3514.90.4.597} {On racial
  diversity and group decision making: identifying multiple effects of racial
  composition on jury deliberations.}
\newblock \emph{Journal of personality and social psychology}, 90(4):597.

\bibitem[{Stelmakh et~al.(2023{\natexlab{a}})Stelmakh, Rastogi, Liu, Chawla,
  Echenique, and Shah}]{stelmakh2023citeseeing}
Ivan Stelmakh, Charvi Rastogi, Ryan Liu, Shuchi Chawla, Federico Echenique, and
  Nihar~B Shah. 2023{\natexlab{a}}.
\newblock \href {https://doi.org/10.1371/journal.pone.0283980} {Cite-seeing and
  reviewing: A study on citation bias in peer review}.
\newblock \emph{PLoS ONE}, 18(7):e0283980.

\bibitem[{Stelmakh et~al.(2023{\natexlab{b}})Stelmakh, Rastogi, Shah, Singh,
  and Daum{\'e}~III}]{stelmakh2020herding}
Ivan Stelmakh, Charvi Rastogi, Nihar~B Shah, Aarti Singh, and Hal
  Daum{\'e}~III. 2023{\natexlab{b}}.
\newblock \href {https://doi.org/10.1371/journal.pone.0287443} {A large scale
  randomized controlled trial on herding in peer-review discussions}.
\newblock \emph{PLoS ONE}, 18(7):e0287443.

\bibitem[{Teplitskiy et~al.(2019)Teplitskiy, Ranu, Gray, Menietti, Guinan, and
  Lakhani}]{teplitskiy2019experts}
Misha Teplitskiy, Hardeep Ranu, Gary~S Gray, Michael Menietti, Eva Guinan, and
  Karim~R Lakhani. 2019.
\newblock \href
  {https://www.hbs.edu/ris/Publication%20Files/19-107_06115731-d0ae-4a11-ab1d-ecaec2118921.pdf}
  {Do experts listen to other experts? field experimental evidence from
  scientific peer review}.

\bibitem[{Tomkins et~al.(2017)Tomkins, Zhang, and
  Heavlin}]{tomkins2017doubleblind}
Andrew Tomkins, Min Zhang, and William~D. Heavlin. 2017.
\newblock \href {https://doi.org/10.1073/pnas.1707323114} {Reviewer bias in
  single- versus double-blind peer review}.
\newblock \emph{Proceedings of the National Academy of Sciences},
  114(48):12708--12713.

\bibitem[{Ware(2008)}]{ware2008peer}
Mark Ware. 2008.
\newblock \emph{Peer review: benefits, perceptions and alternatives}.
\newblock Citeseer.

\bibitem[{Xiong and Litman(2011)}]{Xiong2011Helpfulness}
Wenting Xiong and Diane Litman. 2011.
\newblock \href {https://aclanthology.org/P11-2088} {Automatically predicting
  peer-review helpfulness}.
\newblock In \emph{Proceedings of the 49th Annual Meeting of the Association
  for Computational Linguistics: Human Language Technologies}, pages 502--507,
  Portland, Oregon, USA. Association for Computational Linguistics.

\bibitem[{Yuan et~al.(2022)Yuan, Liu, and Neubig}]{yuan2022automated}
Weizhe Yuan, Pengfei Liu, and Graham Neubig. 2022.
\newblock \href {https://doi.org/10.1613/jair.1.12862} {Can we automate
  scientific reviewing?}
\newblock \emph{Journal of Artificial Intelligence Research}, 75:171--212.

\bibitem[{Zumel~Dumlao and Teplitskiy(2023)}]{dumlao2023geographical}
James~M. Zumel~Dumlao and Misha Teplitskiy. 2023.
\newblock \href {https://doi.org/10.31235/osf.io/754e3} {The effect of reviewer
  geographical diversity on evaluations is reduced by anonymizing submissions}.
\newblock \emph{SocArXiv}.

\end{thebibliography}

\appendix
\section{Formulas}\label{appendix:formulas}

\paragraph{Lexical coverage}
As detailed in \autoref{subsec:coverage}, we measure lexical coverage of paper as the number of n-grams in the submitted paper's abstract that appears in at least one review. More specifically,  if $R_1$ and $R_2$ represent the review text of the pair of reviewers and $S$ represents the submitted paper's abstract, then the lexical coverage is calculated as
\begin{equation}
    \textsc{Lex-Cov}(R_1, R_2, S) = \sum_{n\in \{1, 2, 3\}}\frac{|(\operatorname{ngram}(R;n)) \cap \operatorname{ngram}(S;n)|}{|\operatorname{ngram}(S;n)|},
\end{equation}
where $R$ is the concatenation of the two reviews. The function $\operatorname{ngram}(x; n)$ calculates the n-gram tokens in the text $x$ for $n=\{1,2,3\}$, and $|\mathord{\cdot}|$ calculates the number of unique tokens in the set. 

\paragraph{Semantic coverage}
As detailed in \autoref{subsec:coverage}, we measure semantic coverage of paper as the aggregate similarity of the sentences in the submitted paper abstract, $S$, with the most similar review sentence across the pair of reviews $R_1$ and $R_2$. Specifically, let $R$ be the concatenation of the two reviews, then the semantic coverage is calculated as 
\begin{equation}\label{eq:sentcov}
    \textsc{Sem-Cov}(R_1, R_2, S) = \sum_{s_i \in S} \max\limits_{s_j \in R} \big(\textbf{e}(s_i) \cdot  \textbf{e}(s_j)\big),
\end{equation}
where $s_i$ and $s_j$ represent sentences in the abstract and reviews, respectively and $S$ $\textbf{e}(s) \in \mathbb{R}^{d}$ is the sentence-BERT vector representation of the sentence $s$, where $d=768$.  

\paragraph{Lexical redundancy}
As mentioned in \autoref{subsec:redundancy}, lexical redundancy measures the overlap between the n-grams in the two reviews. Specifically, if $R_1$ and $R_2$ represent the review text of the pair of reviewers and $\operatorname{ngram}(x; n)$ calculates the n-gram tokens in the text $x$ for $n=\{1,2,3\}$, then lexical redundancy is calculated as
\begin{equation}
    \textsc{Lex-Red}(R_1, R_2) = \sum_{n} |\operatorname{ngram}(R_1; n) \cap \operatorname{ngram}(R_2; n)|
\end{equation}

\paragraph{Semantic redundancy}
As mentioned in \autoref{subsec:redundancy}, semantic redundancy measures the cosine similarity between the high-diemsnional representation of the review sentences across the pair of reviews. Specifically, we calculate semantic redundancy as the aggregate similarity of sentences in one review (\textit{reference}) with the most similar sentence in the other review (\textit{target}). To ensure symmetry in the measure, we consider each review as the reference review in turn. 
For a pair of reviews $R_1$ and $R_2$, semantic redundancy is calculated as 
\begin{equation}\label{eq:pairsentsim}
    \textsc{Sem-Red}(R_1, R_2) = \sum_{s_i \in R_1} \max\limits_{s_j \in R_2} \big( \textbf{e}(s_i) \cdot \textbf{e}(s_j)\big) + \sum_{s_i \in R_2} \max\limits_{s_j \in R_1}\big( \textbf{e}(s_i) \cdot \textbf{e}(s_j)\big),
\end{equation}
where $\textbf{e}(s)$ is the sentence-BERT representation \citep{reimers2019sbert} of the review sentences. 

\paragraph{Weighted semantic redundancy}
As mentioned in \autoref{subsec:redundancy}, we measure weighted semantic redundancy as the overlap in the semantic representation of sentences in the pair of reviews weighted by the similarity between their type classification. To annotate the review type, we use the aspect and argument classifier described in \autoref{subsec:coverage}. These classifiers take a review sentence as input and produce a vector representing the probability of that sentence belonging to the different aspect or argument types, with the dimensionality of the probability vector being the number of aspect or argument types. We denote these review type probability vectors as $\textbf{a}(s)$ (where $s$ is a sentence in a review). 

We weigh the semantic similarity of review sentence $s_i$ (coming from review $R_1$) and review sentence $s_j$ (coming from review $R_2$) with the cosine similarity of the aspect or argument type vectors $\textbf{a}(s_i)$ and $\textbf{a}(s_j)$. More specifically, the weighted semantic redundancy is calculated as 
\begin{equation}\label{eq:pairsentsimarg}
    \textsc{Weighted-Sem-Red}(R_1, R_2) = \sum_{s_i \in R_1} \sum_{s_j \in R_2} \big( \textbf{a}(s_i) \cdot \textbf{a}(s_j) \big) \text{ } \big(\textbf{e}(s_i) \cdot \textbf{e}(s_j)\big). 
\end{equation}
Note that in this formulation, we consider similarities between all pairs of sentences in the two reviews, allowing for a more flexible similarity scoring than the semantic redundancy measure, where we only consider the sentence with the maximum semantic similarity. 

\section{Post-processing}\label{appendix:postprocessing}
We approximate the range of our outcome measures for normalization using publicly available review data from a different machine learning conference---\textit{International Conference on Learning Representations} (ICLR 2019),\footnote{\url{https://iclr.cc/}} which includes publicly available reviews from $1565$ submissions. Being a machine-learning conference, the reviewer pools between ICML (the conference that we do analysis on) and ICLR (the conference we estimate the outcome ranges from) have high overlap, helping us in ensuring that the outcome ranges are comparable. However, using different data for estimating the outcome range helps prevent any inadvertent contamination of the causal graph by introducing signals from reviews of other submitted papers. We do not conduct analysis on this data in our study, as we do not have access to the reviewer assignment information for this conference.

\section{Human annotation}\label{appendix:human}
To validate our proposed lexical and semantic measures, we conducted human annotations. Two graduate students with backgrounds related to the conference venue annotated three reviews for $15$ submitted papers. The annotators identified similar and dissimilar elements in two of the reviews, with the third review (chosen at random) serving as a reference review. To address potential calibration differences, we framed the agreement assessment as a ranking problem: whether a pair of annotators or an annotator and a proposed measure ranked the overlap of the two reviews similarly with respect to the reference review.

Our analysis revealed a $77\%$ agreement between the annotators (p-value=$0.046$). The average agreement of an annotator and a proposed measure is $100\%$ (p-value=$0.00097$) and $80\%$ (p-value=$0.05$) for the semantic redundancy and weighted semantic redundancy measures, respectively, which surpasses the agreement between the annotators themselves. 
However, the lexical redundancy measure exhibited random agreement with the annotators (p-value=$0.623$). Since the semantic measures align with human judgments, our findings (discussed in \autoref{sec:results}), which are consistent for semantic measures, are considered valid.

\section{Non-parametric approach}\label{appendix:non-parametric}
The non-parametric approach comprises of three parts---propensity scoring, propensity score matching, and effect estimation. For propensity scoring, we want to measure the probability of treatment (diversity between reviewers along dimension $d^*$) given the other covariates (that is, reviewers' expertise, reviewers' profiles, diversity between reviewers along other dimensions, and the submission). However, similar to the parametric case, controlling for submission is not directly possible due to the high-dimensional nature of the data. Instead, we estimate propensity scores using other covariates (excluding submissions) and then perform matching using propensity scores within submissions. 

Another potential issue with propensity scoring is the ordering between the profile or expertise features for the pair of reviewers in the logistic regression model. To circumvent this, similar to our parametric analysis, we consider an anchor reviewer within each submitted paper and take a diverse and a non-diverse reviewer paired with the anchor reviewer.  
For the submitted paper $S$, let $(r_1, r_2)$ be a pair of diverse reviewers and $(r_1, r_3)$ be a pair of non-diverse reviewers along a dimension $d^*$. Subsequently, we estimate
\begin{equation}
    P\big(\delta_{d^*}(r_1, r_j) \big| \{\{\mathbf{d}(r_j)\}_{d \in D}, E(r_j, S), \{\delta_d(r_1, r_j)\}_{\substack{d\in D \\ d\neq d^*}} \big),
    \label{eq:propensity}
\end{equation}
where $\delta_{d^*}(r_1, r_2)=1$ and $\delta_{d^*}(r_1, r_3)=-1$. 
We train multiple logistic regression models, one for each dimension of diversity $d^* \in D = $ $\{$organization, geographical location, seniority, topic, and co-authorship$\}$. 

Next, for propensity score matching, we consider diverse and non-diverse pairs of reviewers within each submitted paper. Following the previous notation, for the submission $S$, let $(r_1, r_2)$ be a pair of diverse reviewers and $(r_1, r_3)$ be a pair of non-diverse reviewers along a dimension $d^*$. 
We match these two pairs of reviewers if $|P(\delta_{d^*}(r_1, r_2) | \mathbf{X}(r_1, r_2)) - P(\delta_{d^*}(r_1, r_3) | \mathbf{X}(r_1, r_3))| < 0.1$, where $\mathbf{X}$ represents all the covariates in \autoref{eq:propensity}. We discard the submissions that do not have matched diverse and non-diverse pairs of reviewers. This results in $36$, $174$, $113$, $106$, and $96$ matched data points for organization, geographical location, seniority, topic, and co-authorship dimensions ($d^*$) respectively. Each data point comprises of $(S^{(i)}, r_1^{(i)}, r_2^{(i)}, r_3^{(i)})_{i=1}^n$, such that each of the following condition is satisfied:
\begin{itemize}
    \item $\delta_{d^*}(r_1^{(i)}, r_2^{(i)})=1$,
    \item $\delta_{d^*}(r_1^{(i)}, r_3^{(i)})=-1$, and
    \item $|P(\delta_{d^*}(r_1^{(i)}, r_2^{(i)}) | \mathbf{X}(r_1^{(i)}, r_2^{(i)})) - P(\delta_{d^*}(r_1^{(i)}, r_3^{(i)}) | \mathbf{X}(r_1^{(i)}, r_3^{(i)}))| < 0.1$. 
\end{itemize}
Finally, for effect estimation, we calculate 
\begin{equation*}
    \gamma^* = \frac{1}{n} \sum_{i=1}^n y(r_1^{(i)}, r_2^{(i)}; S^{(i)}) - y(r_1^{(i)}, r_3^{(i)}; S^{(i)}).
\end{equation*}
Lastly, to quantify the significance of the difference in the review redundancy or coverage ($y$), we perform permutation test \citep{fisher1936permutation} using scipy \texttt{stats} library\footnote{https://docs.scipy.org/doc/scipy/reference/stats.html} with $10,000$ permutations. 

\section{Topic Representation}\label{appendix:topic}
As discussed in the \autoref{subsec:confounders}, we estimate a reviewer's topic by running a topic modeling method on the abstracts of the reviewer's previously authored papers available on their Google Scholar profile. We use the Latent Dirichlet Allocation (LDA) method \citep{blei2001lda}, a generative probabilistic method used for topic modeling. LDA assumed that each document is a mixture of topics and each topic is a distribution over words. 
For instance, a topic in a machine learning venue, such as ICML, could be reinforcement learning or graphical networks. A reinforcement learning topic may assign high probabilities to words like ``agent'', ``policy'', etc. LDA learns the document-topic distribution and topic-word distribution by iteratively updating these distributions to maximize the likelihood of the observed data (in this case, the collection of reviewer abstracts). The number of topics in the LDA model is a hyperparameter, which we determine by optimizing the topic coherence score \citep{roder2015topiccoherence}, resulting in 10 distinct topics.  

Examples of topics learned by LDA on our dataset are as follows.  Note that the LDA topic only yields the distribution of words in each topic, not the corresponding names. We name the topics listed here manually for exposition. 

\begin{enumerate}
    \item \textit{Computer vision}: image, object, method, video, segmentation, feature
    \item \textit{Reinforcement learning}: agent, policy, learning, reinforcement, state, action
    \item \textit{Recommendation systems}: recommendation, rank, preference, feedback, user, item
    \item \textit{Natural language processing}: language, task, text, query, information, learn
    \item \textit{Systems and security}: system, application, design, time, signal, attack
\end{enumerate}

Using this topic model, we estimate a reviewer’s topic profile by computing the topic distribution of the document containing the abstracts of their previously authored papers. This topic vector serves as a representation of the reviewer’s areas of expertise, based on their past publications. 

\end{document}